\documentclass[conference]{IEEEtran}

\usepackage{cite}
\usepackage{amsmath,amssymb,amsfonts}
\usepackage{mathtools}
\usepackage{algorithmic}
\usepackage{graphicx}
\usepackage{subcaption}
\usepackage{textcomp}
\usepackage{braket}
\usepackage{tabularx, booktabs, makecell, caption}
\usepackage{siunitx}
\usepackage{nicefrac}
\usepackage{comment}
\usepackage{url}
\def\BibTeX{{\rm B\kern-.05em{\sc i\kern-.025em b}\kern-.08em
    T\kern-.1667em\lower.7ex\hbox{E}\kern-.125emX}}

\newcommand*\diff{\mathop{}\!\mathrm{d}}
\begin{document}

\title{Machine Failure Detection Based on \\ Projected Quantum Models}

\author{
\IEEEauthorblockN{
Larry Bowden\IEEEauthorrefmark{1},
Qi Chu\IEEEauthorrefmark{1},
Bernard Cena\IEEEauthorrefmark{1},
Kentaro Ohno\IEEEauthorrefmark{2},
Bob Parney\IEEEauthorrefmark{3},
Deepak Sharma\IEEEauthorrefmark{4},
Mitsuharu Takeori\IEEEauthorrefmark{2}
}
\IEEEauthorblockA{\IEEEauthorrefmark{1}Digital, Woodside Energy, Perth, Australia}
\IEEEauthorblockA{\IEEEauthorrefmark{2}Quantum, IBM Research, Tokyo, Japan}
\IEEEauthorblockA{\IEEEauthorrefmark{3}Quantum, IBM Research, New York, USA}
\IEEEauthorblockA{\IEEEauthorrefmark{4}Quantum, IBM Research, Melbourne, Australia}
}

\maketitle

\begin{abstract}
Detecting machine failures promptly is of utmost importance in industry for maintaining efficiency and minimizing downtime. This paper introduces a failure detection algorithm based on quantum computing and a statistical change-point detection approach.
Our method leverages the potential of projected quantum feature maps to enhance the precision of anomaly detection in machine monitoring systems.
We empirically validate our approach on benchmark multi-dimensional time series datasets as well as on a real-world dataset comprising IoT sensor readings from operational machines, ensuring the practical relevance of our study. 
The algorithm was executed on IBM’s 133-qubit Heron quantum processor, demonstrating the feasibility of integrating quantum computing into industrial maintenance procedures.
The presented results underscore the effectiveness of our quantum-based failure detection system, showcasing its capability to accurately identify anomalies in noisy time series data.
This work not only highlights the potential of quantum computing in industrial diagnostics but also paves the way for more sophisticated quantum algorithms in the realm of predictive maintenance.
\end{abstract}


\maketitle

\section{Introduction}

\begin{figure*}
    \centering
    \includegraphics[width=0.9\linewidth]{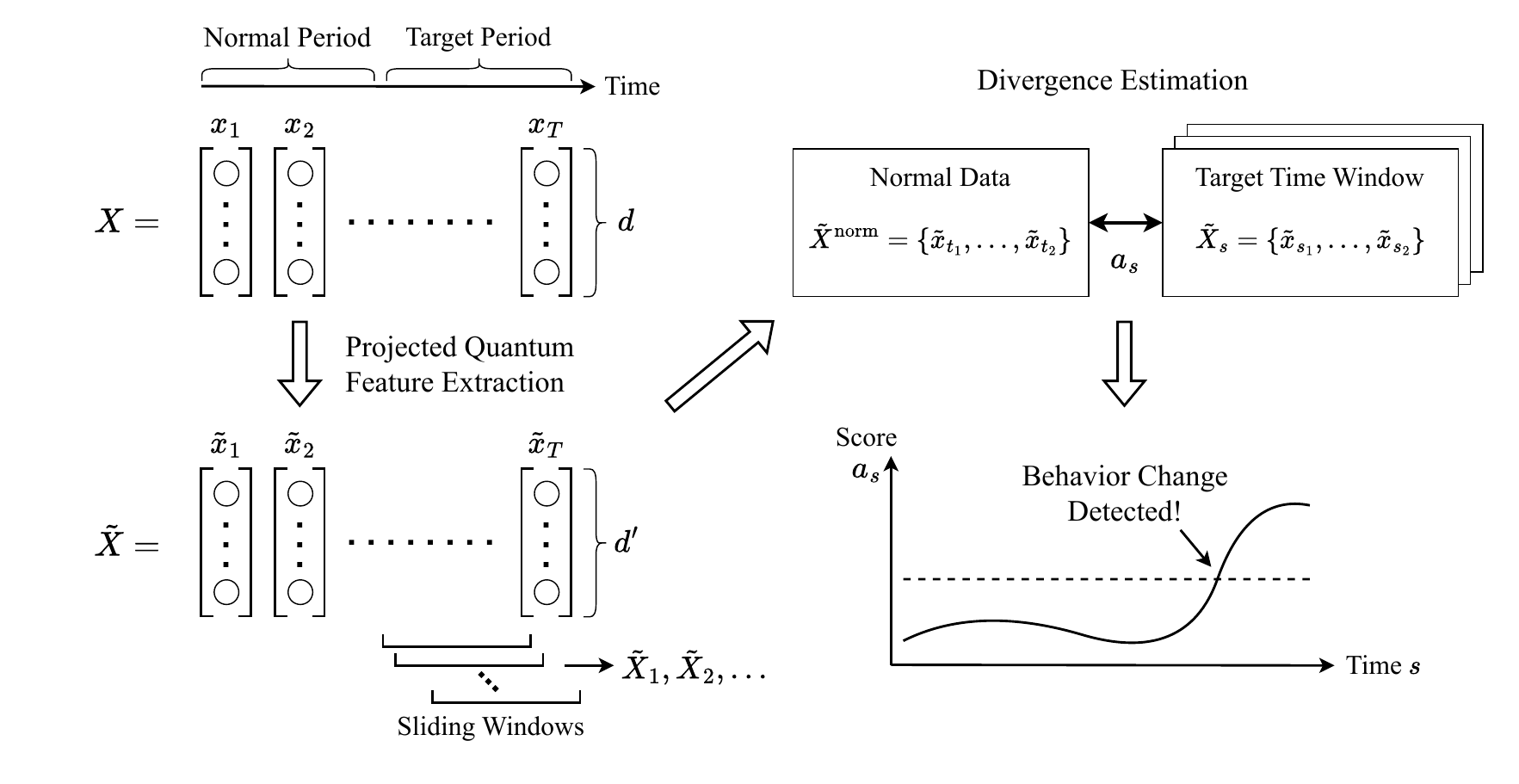}
    \caption{Overview of the proposed framework for machine failure detection based on projected quantum feature extraction and statistical change-point detection approach. Original feature vectors obtained from sensors are transformed to projected quantum features of dimension $d'$ that depends on the quantum circuit used for transformation. New feature vectors are used for statistical divergence estimation to measure anomaly score $a_s$. Large $a_s$ implies that a fault has likely occurred on a machine.}
    \label{fig:overview}
\end{figure*}

In the rapidly evolving industrial landscape, the ability to predict and prevent machine failures has become a critical necessity~\cite{davies1998handbook}.
Traditional methods of fault detection, often reliant on periodic inspections or reactive maintenance strategies, fall short in addressing the nuanced patterns indicative of impending failures. 
These limitations result in unplanned downtime, reduced productivity, and escalating maintenance costs. 
Consequently, there is a pressing need for advanced, proactive diagnostic tools capable of identifying and mitigating potential failures before they cause significant disruptions.

Recent advancements in machine learning technology have opened new avenues for enhanced industrial diagnostics~\cite{lei2020applications,surucu2023condition}. 
Powerful data-driven models can learn from historical and real-time data to predict machine behavior and detect anomalies that may signal impending failures.
Despite the progress, however, achieving high precision is still a challenging task, especially for multidimensional time series~\cite{kuncheva2013pca,faber2021watch}.

This paper explores the integration of quantum computing with machine learning to engineer a superior failure detection algorithm tailored for industrial applications, in the light of the recent remarkable progress in the field of quantum computing~\cite{daley2022practical,cerezo2022challenges,huang2023near,kim2023evidence}.
Quantum computers, with their unique computational capabilities, could be utilized to recognize complex patterns that are beyond the reach of classical computing systems~\cite{mujal2023time}.
In this context, we investigate the potential of quantum computing in developing a cutting-edge failure detection algorithm for machine health monitoring.

Our approach combines the theoretical framework of projected quantum models~\cite{huang2021power} with a statistical change-point detection technique~\cite{aminikhanghahi2017survey}, as schematically shown in Fig.~\ref{fig:overview}.
The machine failure detection is formulated as a change-point detection problem that can be approached with statistical methods such as density ratio estimation~\cite{sugiyama2009density,liu2013change}.
Projected quantum models are a class of quantum machine learning models proposed as an alternative to traditional quantum kernel methods~\cite{havlivcek2019supervised}.
The hybrid methodology would enable us to identify subtle anomalies in multidimensional machine operation data, thereby facilitating timely interventions and preventive maintenance. 

To empirically validate the effectiveness of our proposed framework, we evaluate our methodology on two benchmark datasets of multi-dimensional change-point detection.
The observations of the behavior of our proposed quantum model demonstrate the potential of quantum feature transformation to effectively identify change-points in complex time series.
We also employed a real-world dataset derived from IoT sensor readings of operational machines. 
This dataset, encompassing a wide range of operating conditions and failure modes, provides a practical and relevant context for evaluating our method.
The algorithm was implemented and tested on IBM's 133-qubit quantum processor, demonstrating the feasibility and scalability of integrating quantum computing into industrial maintenance procedures. 
The results of our study underscore the potential of quantum-based failure detection systems in improving machine reliability and reducing maintenance costs.
Although the practical application at present carries computational overhead of quantum workload that may not guarantee superior speed over classical computing methods, our findings highlight the transformative potential and broader implications of quantum computing in the realm of predictive maintenance.

The outline of this paper is as follows.
Machine failure detection is formulated and an existing statistical approach is reviewed in Section~\ref{sec:change_point_detection}.
We provide a methodological background for projected quantum models in Section~\ref{sec:pqf}.
Our framework for machine failure detection is introduced in Section~\ref{sec:method}.
The experimental results are presented in Section~\ref{sec:experiments}.
Lastly, Section~\ref{sec:conclusion} concludes this paper.

\section{Statistical Change-Point Detection}\label{sec:change_point_detection}

In this section, we formalize the machine failure detection problem as a statistical change-point detection problem on multi-variate time series.

\subsection{Problem setting}\label{subsec:problem_setting}
First, we describe the generic assumptions for time series data under consideration.

\begin{itemize}[\labelwidth=0pt]
    \item[] \textbf{Multidimensionality}: Each timestamp is associated with multiple data values, e.g., readings from multiple sensors.
    \item[] \textbf{Access to Normal Data}: Data of a certain period of time during which the machines are in normal operation are available.
    \item[] \textbf{Stationarity}: The machines are in a steady state in a wide sense, i.e., the sensor data may contain some periodicity and other general stationary behavior, under normal operations, while the data are subjective to certain amount of noise.
    \item[] \textbf{Persistent Behavior Change due to Failure}: Once a failure occurs, the machine behavior changes persistently until it is fixed.
\end{itemize}

For detecting the behavior change on time series, 
we can make use of statistical change-point detection methods~\cite{aminikhanghahi2017survey}.
Since our task is to detect machine failures as soon as possible after obtaining anomalous data, it is an \emph{online} detection task.
It is also \emph{unsupervised} as only normal data are available.
Statistical approaches to unsupervised change-point detection are typically categorized into parametric and non-parametric methods.
The latter are often more scalable and robust than the former when applied to change-point detection in stationary processes~\cite{aminikhanghahi2017survey}, thus are suitable in our setting.
In these approaches, the target time series is divided into time windows of appropriate size, each of which is considered as a dataset or empirical probability distribution.
Time windows containing timestamps after the anomaly occurrence will be statistically discriminated from the normal distribution, thereby enabling detecting the behavior change.

Let $X = \{x_t\}_{t=1,\ldots, T}$ denote the given target and normal time series,
where $x_t$ are $d$-dimensional vectors for each timestamp $t$.
We assume a sub dataset $X^\mathrm{norm} = \{x_t\}_{t_1\le t \le t_2}$ on an interval $[t_1, t_2]$ is a normal dataset, i.e., does not contain anomalies.
Given window length $L>0$ and sliding width $w>0$, time windows are formally defined as 
\begin{align}\label{eq:time_window}
    X_s = \{x_t\}_{t=sw, \ldots, sw+L}, s=1,2,\ldots.
\end{align}
With this notation, the task is to measure the statistical divergence $a_s$ between distributions of $X^\mathrm{norm}$ and $X_s$ for each $s$, which is viewed as a score representing how likely an anomaly or change has occurred.

Despite huge technical advances in change-point detection algorithms, highly precise detection is still challenging, especially for multidimensional data~\cite{kuncheva2013pca,faber2021watch}.
Among the existing methods, \emph{density ratio estimation}~\cite{kawahara2009change,liu2013change} is empirically the most powerful non-parametric approach to change-point detection~\cite{aminikhanghahi2017survey}.
Therefore, we incorporate quantum modeling into this approach to enhance its accuracy.

\subsection{Computing Divergence via Density Ratio Estimation}

We briefly review the density ratio estimation approach to computing statistical divergence between normal and target sample sets.

Let $p(x)$ and $p'(x)$ be density functions representing probability distributions on a domain in $\mathbb{R}^d$.
Density ratio estimation is to estimate the ratio function $r(x) \coloneqq p(x)/p'(x)$ of the density functions using sample sets for these distributions.
The ratio $r(x)$ can be used to calculate the divergence of two probability distributions.
For example, the 
Pearson divergence $\operatorname{PE}(p \parallel p')$ is written as
\begin{align}
    \operatorname{PE}(p \parallel p') &= \int (r(x)-1)^2 p'(x) \diff x \\
    &= \frac{1}{2}\int r(x) p(x) \diff x - \frac{1}{2}.
\end{align}
The point in directly estimating the ratio $p(x)/p'(x)$ instead of estimating $p$ and $p'$ first and taking the quotient of them is to avoid the accumulation of approximation errors~\cite{sugiyama2009density}.

Algorithms for density ratio estimation consists of two steps: modeling the ratio function with a suitable function and solving a minimization problem to fit the model to the actual density ratio function.
In this work, we employ a method called unconstrained least-squares importance fitting (uLSIF)~\cite{kawahara2009change} since the minimization can be solved analytically, thereby offering a fast and stable computational solution~\cite{sugiyama2009density}.
Let $\mathcal X = \{x_1, \ldots, x_n\}$ and $\mathcal X' = \{x'_1, \ldots, x'_{n'}\}$ denote sample sets from $p(x)$ and $p'(x)$, respectively. 
In uLSIF, the ratio function is modeled by the following linear function:
\begin{align}
    g(x; \alpha) \coloneqq \sum_{j=1}^m \alpha_j \eta_j(x),
\end{align}
where 
$\eta_j(x)$ is a basis function.
Typically, $\eta_j(x)$ is defined as follows using a subsample $x_{i_j} \in \mathcal{X}$ and a Gaussian radial basis function (RBF) kernel $k:\mathbb{R}^d\times \mathbb{R}^d \to \mathbb{R}$ with a scale parameter $l > 0$:
\begin{align}\label{eq:basis_function}
    \eta_j(x) \coloneqq k(x,x_{i_j}) = \exp{\left(- \frac{\lVert x-x_{i_j} \rVert^2}{2l^2} \right)}.
\end{align}
The parameters $\alpha = (\alpha_1, \ldots, \alpha_m)$ are determined so that the following squared error $J$ is minimized:
\begin{align}
    J(\alpha) =&\ \frac{1}{2} \int \left(g(x; \alpha) - \frac{p(x)}{p'(x)} \right)^2 p'(x) \diff x \\
    =&\ \frac{1}{2} \int g(x; \alpha)^2 p'(x) \diff x - \int g(x; \alpha)p(x) \diff x \notag\\ 
    &+ \frac{1}{2} \int \frac{p(x)^2}{p'(x)} \diff x.
\end{align}
Ignoring the last term, which is constant, and approximating the expectation by empirical averages, the following objective is obtained:
\begin{align}
    \hat{J}(\alpha) &= \frac{1}{2n'} \sum_{i=1}^{n'} g(x'_i; \alpha)^2 - \frac{1}{n} \sum_{i=1}^n g(x_i; \alpha) \\
    &= \frac{1}{2} \sum_{j,j'=1}^m \alpha_j \alpha_{j'} \hat{H}_{j,j'} - \sum_{j=1}^m \alpha_j \hat{h}_j,
\end{align}
where
\begin{align}
    \hat{H}_{j,j'} = \frac{1}{n'}\sum_{i=1}^{n'} \eta_{j}(x'_i) \eta_{j'}(x'_i) ,\quad
    \hat{h}_j = \frac{1}{n}\sum_{i=1}^{n} \eta_j (x_i) .
\end{align}
Adding a regularization term $\lambda \lVert \alpha \rVert^2$ with a positive number $\lambda > 0$ to $\hat J(\alpha)$, a minimizer is analytically derived as
\begin{align} \label{eq:alpha_solution}
    \hat{\alpha} \coloneqq (\hat H + \lambda I_m )^{-1} \hat h.
\end{align}
To assure the positivity of the model function $g(x; \hat \alpha)$, we truncate the negative part of $\hat \alpha$:
\begin{align}
    \hat{\alpha}_j \leftarrow \max(0, \hat{\alpha}_j) \ \mathrm{for} \ j=1,\ldots, m.
\end{align}
Using the obtained function $\hat g(x) \coloneqq g(x; \hat \alpha)$, we can estimate the Pearson divergence $\operatorname{PE}(p \parallel p')$ between $p$ and $p'$ as
\begin{align}
    \operatorname{\widehat{PE}}(\mathcal{X}, \mathcal{X}') \coloneqq \frac{1}{2n} \sum_{i=1}^n \hat{g}(x_i) - \frac{1}{2}.
\end{align}
The overall complexity of the algorithm is $O(m^3+n+n')$.

\section{Projected Quantum Models}\label{sec:pqf}

In this section, we review the 
projected quantum model~\cite{huang2021power}, which is the core of our method.
The projected quantum model is invented based on the careful analysis of a necessary condition for potential quantum advantage in quantum machine learning, working around a critical issue in traditional quantum kernel methods.
Below, we show a specific form of projected quantum models, referring to the original paper~\cite{huang2021power} for the analysis results.

Consider a parametrized quantum circuit representing a unitary operator $U(\theta)$ with a parameter vector $\theta$.
Let $n_q$ be the number of qubits.
We assume $\theta$ is $d$-dimensional, that is, has the same dimension as a data point $x \in \mathbb{R}^d$.
For each data point $x$, the quantum circuit maps the initial state $|\psi\rangle$ to $|\phi(x)\rangle \coloneqq U(x)|\psi\rangle$. 
Let $\rho(x)$ denote a density matrix corresponding to $|\phi(x)\rangle$.
In projected quantum models, we take projection of $\rho(x)$ rather than using $\rho(x)$ itself to represent a quantum feature of $x$.
In this study, we employ one-particle reduced density matrix (1-RDM) $\rho_k(x), k=1,\ldots,n_q$ obtained by projecting $\rho(x)$ to each qubits:
\begin{align}
    \rho_k(x) \coloneqq \operatorname{Tr}_{j\ne k} [\rho(x)].
\end{align}
While the projection apparently loses information on entanglement of quantum states, it can actually help improving prediction accuracy of machine learning models~\cite{huang2021power}.
On the basis of this fact, the main idea in this work is to build a detection model based on the 1-RDM features instead of the original feature vectors.
In practice, it is useful to represent $\rho_k(x)$ in the Pauli basis 
$(\sigma_0, \sigma_1, \sigma_2, \sigma_3) = (I, X, Y, Z)$ as
\begin{align}
    \rho_k(x) = \sum_{i=0}^3 c_{k,i} \sigma_i
\end{align}
with real coefficients $c_{k,i}, \ i=0,1,2,3$.
Note that each coefficient satisfies
\begin{align}
    c_{k,i} = \frac{1}{2} \operatorname{Tr}(\rho_k \sigma_i),
\end{align}
resulting in $c_{k,0} = 1/2$.
The rest coefficients $c_{k,1}, c_{k,2}, c_{k,3}$ can be computed on a gate-based quantum computer by taking expected values of observables 
\begin{align}
    I^{\otimes n_q-k} \otimes \sigma_i \otimes I^{\otimes k-1} , \quad 
    i = 1, 2, 3
\end{align}
for the quantum state $|\phi(x)\rangle$.
By concatenating them over $k=1,\ldots,n_q$, we obtain a new feature vector
\begin{align}
    \tilde{x} \coloneqq (c_{1,1}, c_{1,2}, c_{1,3}, c_{2,1}, \ldots, c_{n_q,3}) \in \mathbb{R}^{3n}
\end{align}
representing 1-RDM of all qubits.
We call $\tilde{x}$ the projected quantum feature of $x$. 
By appropriately choosing the quantum circuit for $U(\theta)$, computing projected quantum features can be neither classically simulatable nor learnable by classical machine learning models~\cite{huang2021power}.

\section{Proposed Method}\label{sec:method}

Based on notions introduced in the previous sections, we present a machine failure detection workflow utilizing quantum computers.

Our method consists of the following procedures, as described in Fig.~\ref{fig:overview}.
Let $X = \{x_t\}_{t=1,\ldots, T}$ and $X^\mathrm{norm} = \{x_t\}_{t_1\le t \le t_2}$ be given target and normal time series.
First, we convert the original time series data $X$ into a sequence of 1-RDM projected quantum features $\tilde X = \{\tilde x_t\}$ by running a certain quantum circuit $U(x_t)$ on a quantum computer for each $t=1,\ldots, T$.
Let $\tilde X^\mathrm{norm} = \{\tilde x_t\}_{t_1\le t \le t_2}$ denote the sequence of normal data converted into projected quantum features.
Then, we create sliding time windows $\tilde X_s, s=1,2,\ldots$ from the new time series $\tilde X$ following the same way in Section~\ref{subsec:problem_setting}.
The anomaly score $a_s$ for each time window $\tilde X_s$ is computed as the estimated Pearson divergence 
\begin{align}
    a_s = \operatorname{\widehat{PE}}(\tilde X^\mathrm{norm}, \tilde X_s)
\end{align}
by running uLSIF for two sample sets $\tilde X^\mathrm{norm}, \tilde X_s$ on a classical computer.
Timestamp $s$ having a large value of $a_s$ is viewed as the time after the anomaly or machine failure occurred.
This detection is done by thresholding $a_s$ and the threshold might be adjusted during the machine operation.

\section{Experiments}\label{sec:experiments}

We conduct numerical experiments to validate the proposed method.
First, we investigate the effect of quantum feature transform in the change-point detection setting using simulation of quantum circuits. 
Then, we test our approach to failure detection on actual industrial time series data using quantum hardware as well as a simulator.
On each experiment, our method is compared with the uLSIF method without the use of quantum feature transform to observe how quantum modeling changes the behavior of the algorithm.

\subsection{Experiment Setup}

We use a quantum circuit based on 1-dimensional Heisenberg Hamiltonian, inspired by earlier work \cite{huang2021power,wiersema2020exploring}.
The choice is motivated by the expectation that this circuit is hard to simulate and resistant against `dequantization'~\cite{shin2024dequantizing} unlike other common circuit proposed in previous research~\cite{sim2019expressibility}.
Additionally, its use aligns with recent simulation efforts in quantum chemistry and material science~\cite{kumaran2025quantum}, signifying potential for a clear separation from classical computing methods.
For another choice of the circuit, see Appendix~\ref{app:two-local}.

\begin{figure}[t]
\centering
    \includegraphics[width=\linewidth]{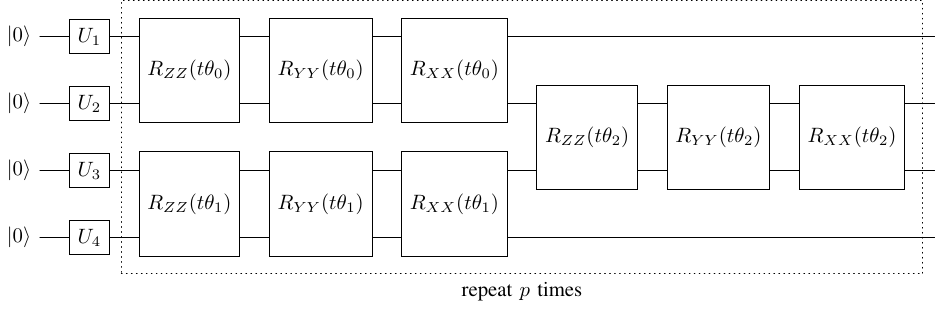}
    \caption{Schematic diagram of the quantum circuit in our experiment for $n_q=4$. $U_1, \ldots,U_4$ are Haar-random unitary gates with a fixed random seed and $R_{XX}, R_{YY}, R_{ZZ}$ are 2-qubit rotation gates with respect to $XX, YY, ZZ$, respectively. $t$ is set to a fixed constant and $\theta$ represents feature vectors.}
    \label{fig:heisenberg_circuit}
\end{figure}

We describe the circuit used in our experiments.
For a system of $n_q$ qubits, we define
\begin{align}
    H^\mathrm{ even}_{\sigma \sigma} = \sum_{i=1}^{\lfloor n_q/2 \rfloor} \theta_{2i-1} \sigma_{2i-1}\sigma_{2i}, \quad
    H^\mathrm{ odd}_{\sigma \sigma} = \sum_{i=1}^{\lfloor n_q/2 \rfloor} \theta_{2i} \sigma_{2i}\sigma_{2i+1}
    \notag
\end{align}
for each Pauli operator $\sigma \in \{X, Y, Z\}$.
Then, the quantum circuit is defined by
\begin{align}
    U(\theta) = &\left( G(H^\mathrm{ odd}_{XX}) G(H^\mathrm{ odd}_{YY}) G(H^\mathrm{ odd}_{ZZ}) \right. \notag\\
     &\left. G(H^\mathrm{ even}_{XX}) G(H^\mathrm{ even}_{YY}) G(H^\mathrm{ even}_{ZZ}) \right)^p,
\end{align}
where $G(H) = \exp(-itH)$ with some constant $t>0$.
The initial state $|\psi\rangle = \otimes_i|\psi_i\rangle$ is defined as a separable state consisting of Haar-random single-qubit quantum state $|\psi_i\rangle$ (the random state is fixed over all data for consistency).
We set $p=1$ so that the circuit is not too deep for obtaining meaningful results from hardware.
We fix the value of $t$ to 0.5 on the basis of a preliminary experiment where we test $t\in [0.1,2]$ on the synthetic data below and take the value yielding the most stable outputs.
The number of qubits is $n_q = d+1$ where $d$ is the feature dimension so that the number of parameters is matched to the feature dimension of the data.
The circuit diagram for $n_q = 4$ is shown in Fig.~\ref{fig:heisenberg_circuit}.
The data point $x$ is encoded as $\theta = \arctan(x)$ to bound the encoding angle to $[-\pi/2, \pi/2]$.
For datasets used in our experiments, $n_q$ is small enough to simulate the quantum circuit exactly.
Therefore, we compute 1-RDM features with a state vector simulator using Qiskit 1.2.0~\cite{Qiskit}.

\subsection{Change-Point Detection Experiments}

We evaluate how the quantum feature transform changes the behavior of detection algorithms on existing multi-dimensional change-point detection datasets.
To qualitatively observe the effect of the quantum model, we deliberately choose small-scale datasets, allowing clear visualization and exact simulation of quantum circuits.
Unlike the failure detection problem having only one change-point, the time series in the datasets have multiple change-points representing transitions between possible states.
In this problem setting, normal data do not exist and instead, each target time window is compared with an adjacent time window that also slides along the time axis.
Specifically, by setting $w=1$ in Eq.~(\ref{eq:time_window}), we compute the divergence between $X_s$ (or $\tilde X_s$) and $X_{s-L}$ (or $\tilde X_{s-L}$) which we call the \emph{change score}.
Change-points are detected at points where the score takes a peak.

For the parameters in uLSIF, the regularization coefficient is fixed to $\lambda=0.1$, as we found that varying the value does not significantly change the output. 
The stability of the uLSIF method heavily depends on the scale parameter $l>0$ in Eq.~(\ref{eq:basis_function}).
Therefore, we vary $l$ for a certain range and report the best value giving the smallest noisy deviation.

\begin{figure}
\begin{minipage}[b]{0.5\linewidth}
    \centering
    \includegraphics[width=\linewidth]{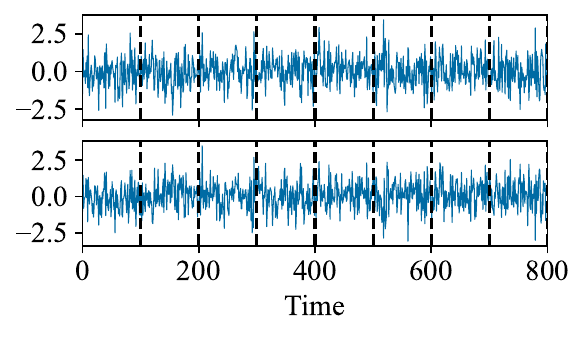}
    \subcaption{Data values.}
    \label{fig:synthetic_data}
    \par\vskip\floatsep
    \includegraphics[width=\linewidth]{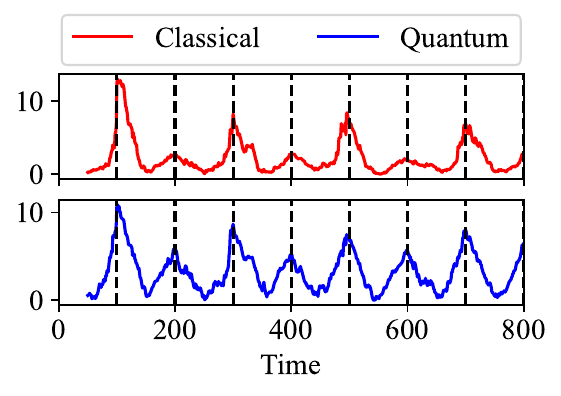}
    \subcaption{Change scores.}
    \label{fig:synthetic_score}
\end{minipage}\hfill
\begin{minipage}[b]{0.5\linewidth}
    \centering
    \includegraphics[width=\linewidth]{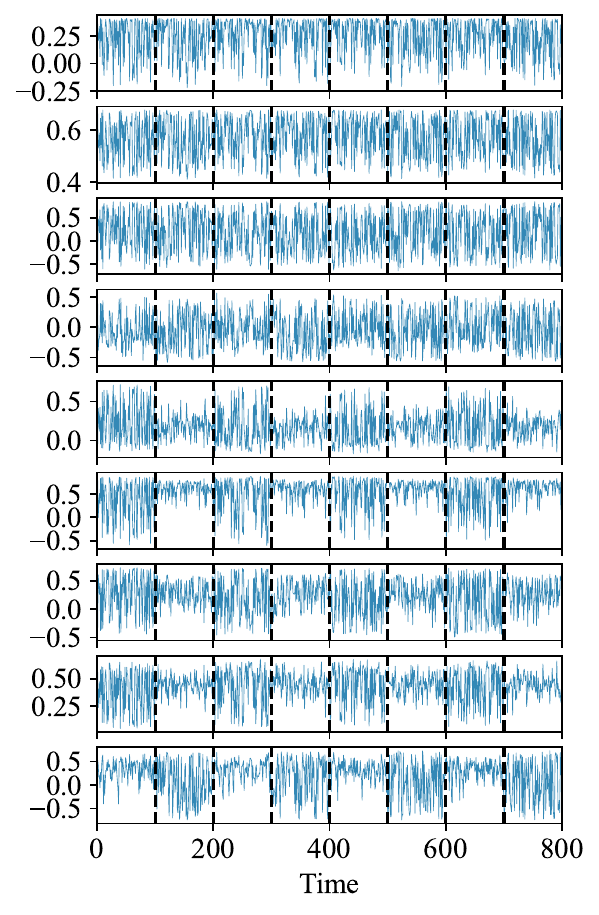}
    \subcaption{1-RDM feature values.}
    \label{fig:synthetic_pqf}
\end{minipage}
\caption{Experiments on 2-dimensional synthetic data. Change-points are represented by black dotted lines.}
\label{fig:synthetic_experiment}
\end{figure}

\subsubsection{Synthetic Data}

We follow the data generation protocol in the existing study~\cite{liu2013change} to produce 2-dimensional synthetic data.
Specifically, we take a sample at each timestamp $t$ from the origin-centered normal distribution with covariance matrix
\begin{align}
    \Sigma = \begin{cases}
        \begin{pmatrix}
            1 & - \frac{4}{5} - \frac{N-2}{500} \\
            - \frac{4}{5} - \frac{N-2}{500} & 1 \\
        \end{pmatrix} & N=1,3,5,\ldots \\
        \begin{pmatrix}
            1 & \frac{4}{5} + \frac{N-2}{500} \\
            \frac{4}{5} + \frac{N-2}{500} & 1 \\
        \end{pmatrix} & N=2,4,6,\ldots
        \end{cases}
\end{align}
for $t\in [100(N-1),100N]$.
Fig.~\ref{fig:synthetic_data} shows the generated data.
Note that the changes are not clearly visible in the data. 

For computing the change score, the window length $L$ is set to 50.
The basis function in Eq.~(\ref{eq:basis_function}) is computed using the first 50 data points, resulting in 50 basis functions.
The scale parameter $l$ in Eq.~(\ref{eq:basis_function}) is varied over $\{0.5, 1, 2, 5\}$, and the reported best value is $l=1$ and $l=5$ for the classical and quantum detection model, respectively.

The change scores obtained by usual uLSIF (`Classical') and uLSIF on 1-RDM feature (`Quantum') are shown in Fig.~\ref{fig:synthetic_score}.
While the classical model already detects the change-points fairly well, we see that the quantum model emphasizes them better by showing clearer peaks.
To investigate the reason of this result, we plot the computed 1-RDM feature vectors in Fig.~\ref{fig:synthetic_pqf}.
Interestingly, it is observed that the behavior change (alternating between two states) in the data are clearly visualized for several dimensions in the 1-RDM feature.
We consider that this separation makes it easier for the model to capture the change-points.

\begin{figure}
\begin{minipage}[b]{0.49\linewidth}
    \centering
    \includegraphics[width=\linewidth]{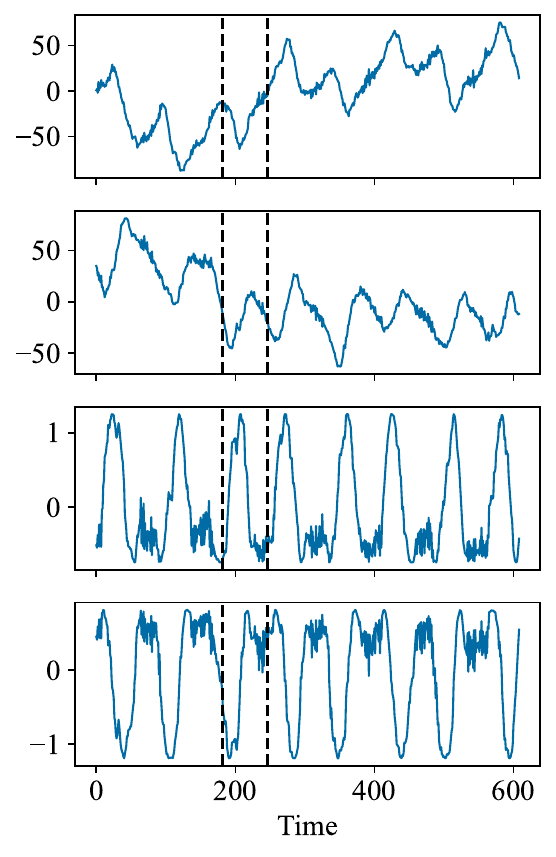}
    \subcaption{Data values.}
    \label{fig:bee_data}
\end{minipage}
\begin{minipage}[b]{0.5\linewidth}
    \centering
    \includegraphics[width=\linewidth]{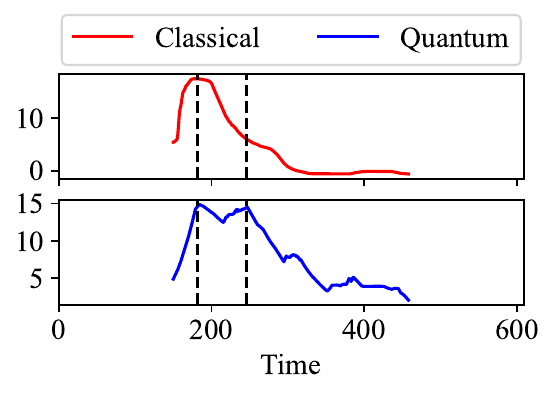}
    \subcaption{Change scores.}
    \label{fig:bee_score}
    \par\vskip\floatsep
    \includegraphics[width=\linewidth]{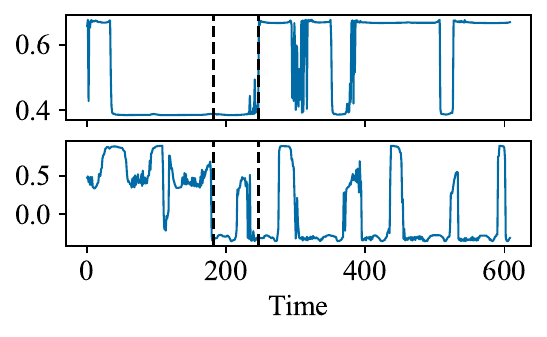}
    \subcaption{1-RDM feature obtained by observables $IIIIY$ and $IZIII$.}
    \label{fig:bee_pqf}
\end{minipage}
\caption{Experiments on bee-waggling data. Change-points are represented by black dotted lines.}
\label{fig:bee_experiment}
\end{figure}

\begin{figure*}[t]
    \begin{subfigure}{.48\textwidth}
        \centering
        \includegraphics[width=0.95\linewidth]{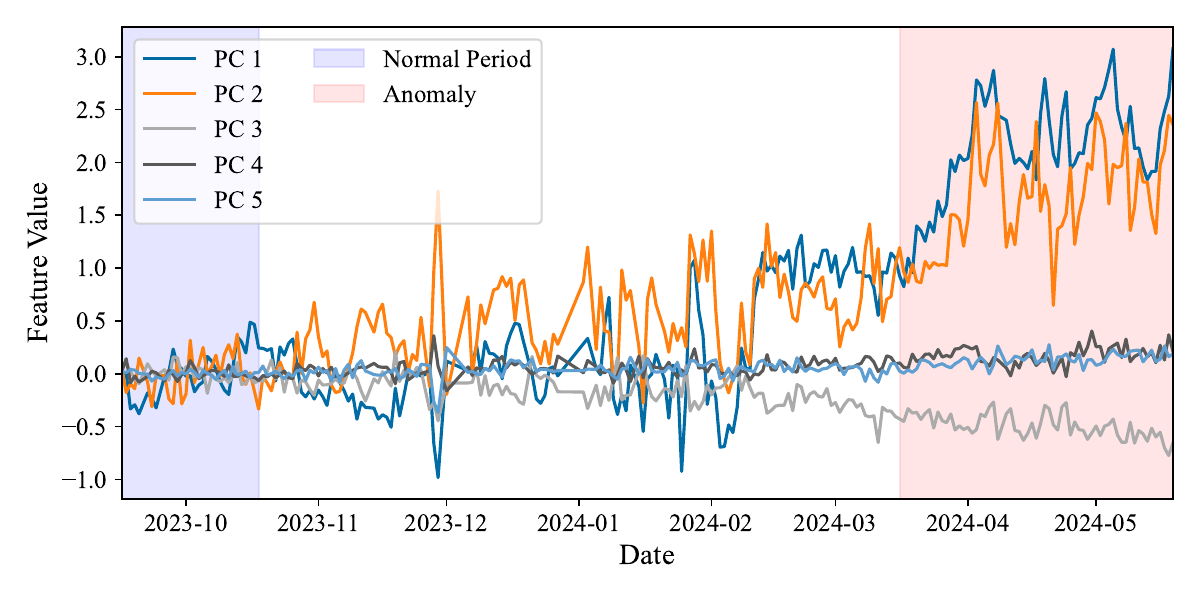}
        \subcaption{Sequence 1.}\label{fig:data_1}
    \end{subfigure}
    \begin{subfigure}{.48\textwidth}
        \centering
        \includegraphics[width=0.95\linewidth]{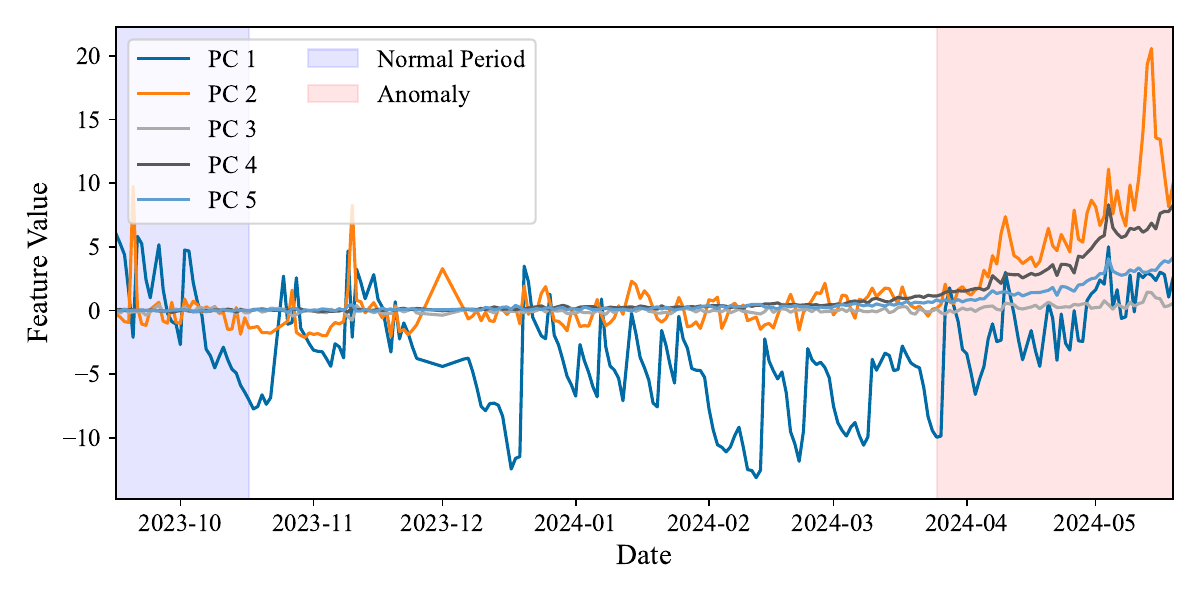}
        \subcaption{Sequence 7.}\label{fig:data_7}
    \end{subfigure}
    \caption{Examples of time series data collected from IoT devices. 
    For visibility, the dimension of plotted feature vectors are reduced from 13 to 5 by using the top five significant principal components (PCs). PC $i$ represents the feature value projected onto the $i$-th principal component. Normal period (blue shaded) is used as reference data distribution to compute the statistical divergence to each target time window. Anomaly period (red shaded) is annotated as ground truth.}
    \label{fig:sequence_data}
\end{figure*}

\subsubsection{Bee Waggle Dance Data}

We take a public time series dataset for change-point detection from the existing study~\cite{van2020evaluation}.
The dataset contains two 2-dimensional time series (\textit{apple}, \textit{run\_log}) and two 4-dimensional time series (\textit{bee\_waggle\_6}, \textit{occupancy}).
We choose \textit{bee\_waggle\_6} data (Fig.~\ref{fig:bee_data}) for our test, since other sequences contains a number of discontinuous points and most change-points can be detected only by monitoring such discontinuity.

The window length is set to 150 and the basis function is computed using the first 150 data points.
The scale parameter $l$ is varied over $\{1, 5, 10, 20\}$ and $l=10$ is found to be the best for both classical and quantum models.

The change scores are shown in Fig.~\ref{fig:bee_score}.
We see that the quantum model detects the two change-points accurately, while the usual uLSIF model is missing the second change-point.
Meanwhile, the quantum score takes several peaks on the low-score regime, possibly interpreted as the false-positive.
Practically, these models could be used selectively taking the difference in the model behavior and the tradeoff between sensitivity and specificity into account.
We also plot 1-RDM feature values of major two dimensions that visually well describe the state transition in Fig.~\ref{fig:bee_pqf}.
Unlike the original data, we observe abrupt changes in 1-RDM feature aligned with change-points, which presumably improves the model accuracy.

\subsection{Machine Failure Detection on Sensor Data}

We validate the proposed method on actual industrial data with quantum device as well as the simulator.

\subsubsection{Dataset Description}

We introduce the real-world sensor data used in this experiment.
The data streams were collected with IoT devices monitoring vibration of fin fan coolers on propane condenser heat exchangers.
A data point was recorded for every 30 minutes.
After preprocessing in the devices, the feature vector at each timestamp is given as 13 dimensional features including acceleration and accumulated velocity characterizing vibration.
Our task is to detect vibration anomalies for fans and motors of fans such as belt damage or fan bearing defect.
Anomaly labels are defined on some data sequences as a period by domain experts, which are used as the ground truth for the evaluation of models.

We used the labeled sequences of a period from 2023-09-15 to 2024-05-20.
The normal period is set to the first 30 days and the rest is treated as a target period, from which we created target time windows of length 7 days with a sliding width of 1 day. 
The sequences including an anomaly in the first 30 days are excluded from the evaluation.
As a result, seven sequences were chosen for evaluation.

The original dataset has around 5000 timestamps in each sequence.
Since computing 1-RDM for this large amount of data points could be expensive on current quantum computers, we averaged the data over each day to create a smaller dataset.
The length of resulting sequences are around 200.
Although the aggregation might change the performance of the change-point detection algorithm, we confirm that it does not change the result significantly in our case, see Appendix~\ref{appendix:experiment_raw_dataset} for details.

\subsubsection{Computational Setup}

We execute the quantum circuit on real quantum hardware in addition to the simulator.
For the hardware experiments, \texttt{ibm\_torino}, a 133-qubit Heron device, is used with TREX~\cite{van2022model} for error mitigation.
The number of shots for estimating 1-RDM of each data point is set to 8192.

For applying uLSIF, the number of the basis functions $\eta_j$ equals to the number of data points in the normal period.
The scale parameter $l>0$ is varied for $l=\{0.1,1,10\}$ and the most stable result is obtained with $l=1$ for the classical model and $l=0.1$ for the quantum model. 

\begin{figure*}[t]
    \begin{subfigure}{.48\textwidth}
        \centering
        \includegraphics[width=0.95\linewidth]{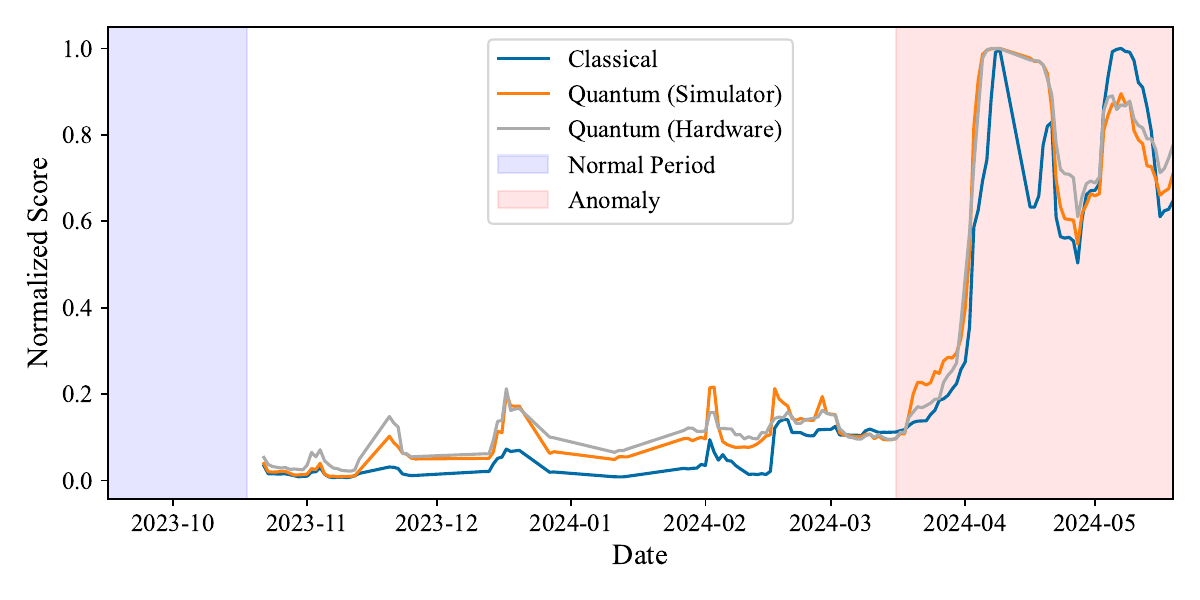}
        \subcaption{Sequence 1.}\label{fig:score_1}
    \end{subfigure}
    \begin{subfigure}{.48\textwidth}
        \centering
        \includegraphics[width=0.95\linewidth]{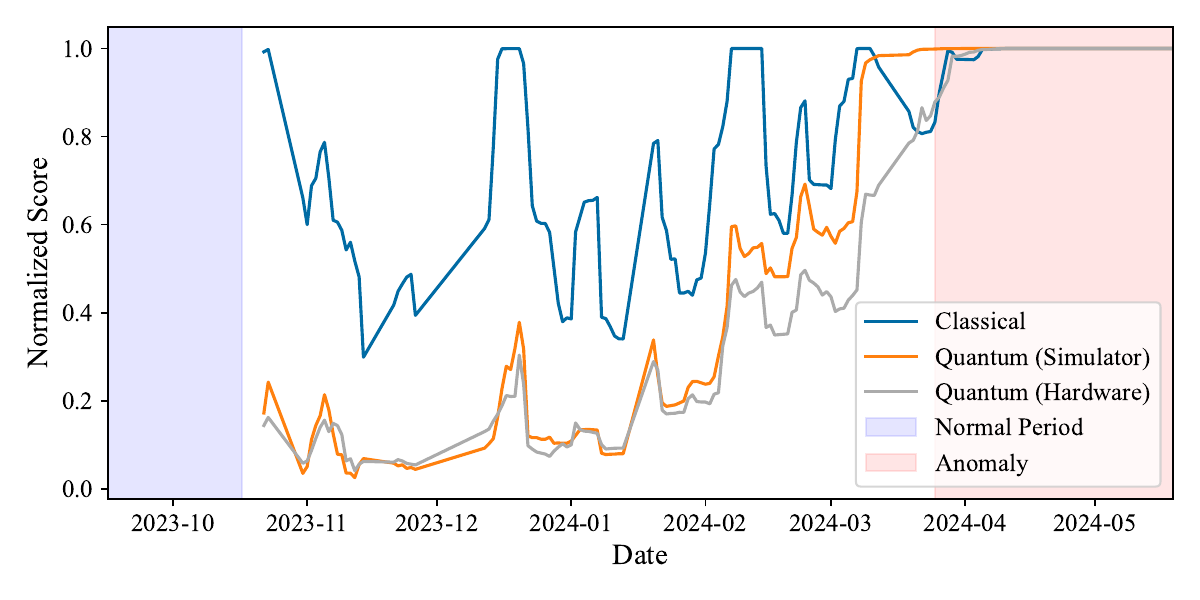}
        \subcaption{Sequence 7.}\label{fig:score_7}
    \end{subfigure}
    \caption{Anomaly score $a_s$ computed by uLSIF methods on raw features and projected quantum features. The scores are divided by the maximum over the whole period for ease of comparison. Values on the x-axis correspond to the end of the time window for which the score is calculated. Therefore, the scores are plotted from 2023-10-21, the last day of the first target time window after the normal period.}
    \label{fig:anomaly_score}
\end{figure*}

\begin{figure*}[t]
    \begin{subfigure}{.48\textwidth}
        \centering
        \includegraphics[width=0.95\linewidth]{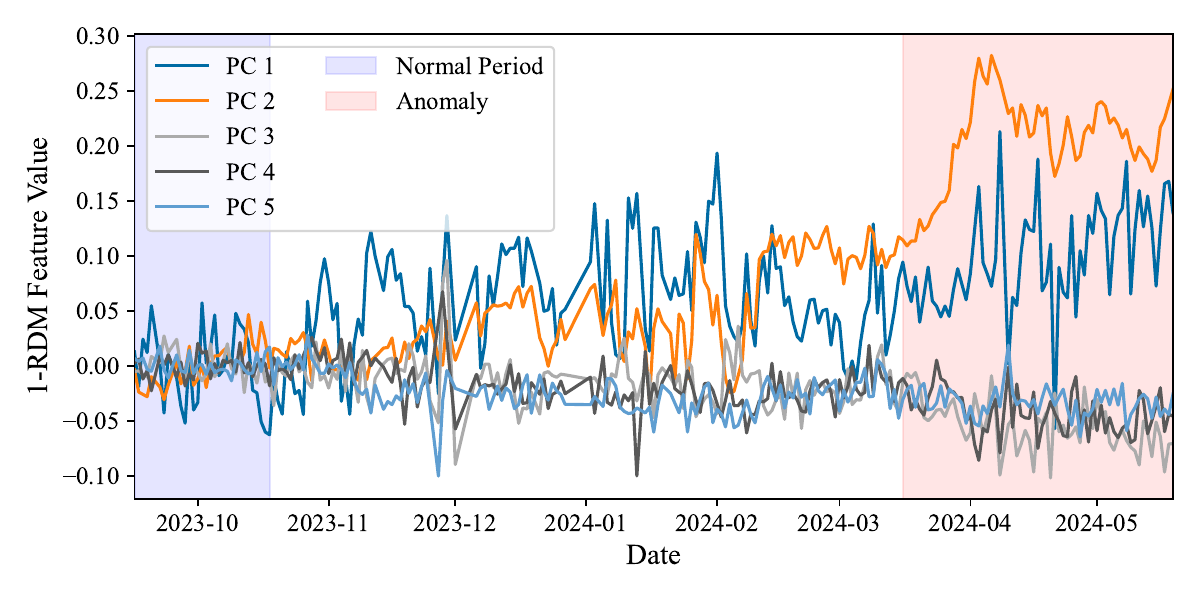}
        \subcaption{Sequence 1.}\label{fig:pqf_1}
    \end{subfigure}
    \begin{subfigure}{.48\textwidth}
        \centering
        \includegraphics[width=0.95\linewidth]{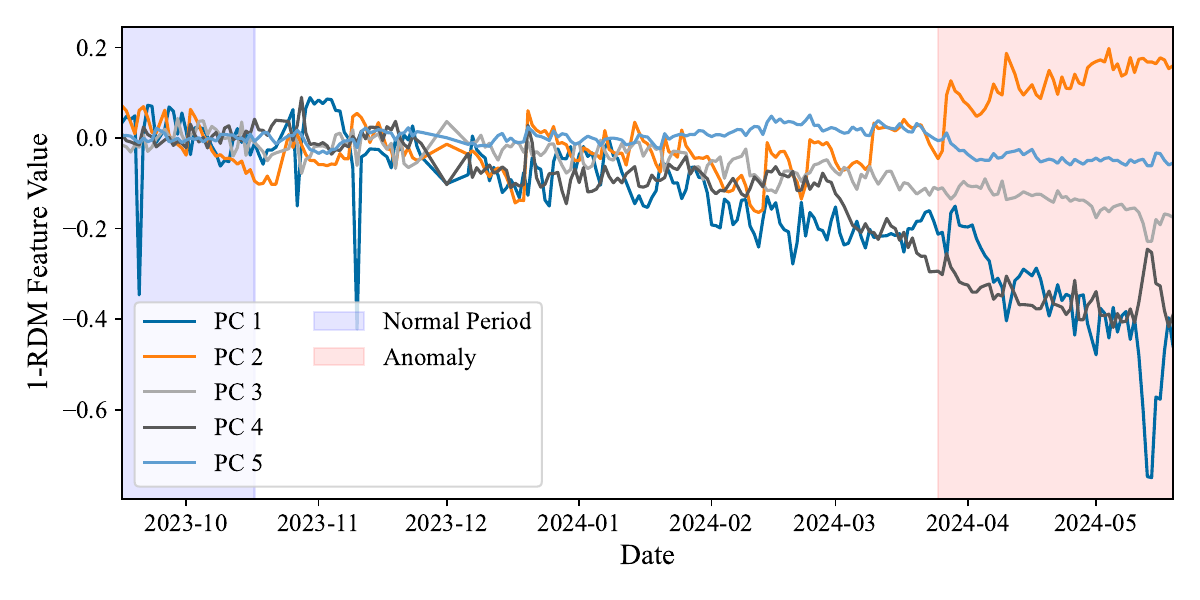}
        \subcaption{Sequence 7.}\label{fig:pqf_7}
    \end{subfigure}
    \caption{
    1-RDM feature vectors plotted with dimensionality reduction by PCA as in Fig.~\ref{fig:sequence_data}.
    }
    \label{fig:pqf}
\end{figure*}

\subsubsection{Results}

First, we make a qualitative observation on the behavior of detection models taking two sequences, Sequence 1 and 7, from the dataset.
The time series are plotted in Fig.~\ref{fig:sequence_data}.
Since plotting 13-dimensional time series could be messy, we reduced the feature dimension to 5 using principal component analysis (PCA) in the figure, where the principal components are computed using the data of the normal period.
Fig.~\ref{fig:anomaly_score} shows the output $a_s$ of the classical and quantum models, which is computed by uLSIF as $a_s = \widehat{\operatorname{PE}}(X^\mathrm{norm}, X_s)$ for classical and $a_s = \operatorname{\widehat{PE}}(\tilde X^\mathrm{norm}, \tilde X_s)$ for quantum, respectively.
On Sequence 1, the classical and quantum models perform similarly: $a_s$ is small before the anomaly period and large on the anomaly period, indicating that both models successfully detect the machine failure.
Additionally, the score obtained by the hardware execution is well aligned to the simulator result.
This shows that the hardware is producing outputs close to theoretical values, and we expect this behavior to hold true even on larger scales that cannot be simulated classically.
Sequence 7 is another instance that involves more noise with respect to the first principal component as shown in Fig.~\ref{fig:data_7}.
On this sequence, the score $a_s$ obtained from the classical model takes large values on most timestamps as shown in Fig.~\ref{fig:score_7}, leading to the false positive prediction.
By contrast, the quantum model outputs large scores only on the anomaly period.
This result indicates that the quantum model is more robust to noise in the data.
To observe this effect more clearly, we plot the obtained 1-RDM feature in Fig.~\ref{fig:pqf}.
From the figure, we confirm that the noisy behavior in the original data of Sequence~7 has indeed disappeared.
On the other hand, on Sequence~1, the transformed data look noisier than the original data over the period before the anomaly, which could be the reason why the quantum anomaly score is higher than the classical one over that period.
These results provide the insights into how quantum models deal with data noise, which is an interesting future direction in quantum machine learning research.

\begin{table}[t]
  \caption{AUC Scores on Real Industrial Time Series.}
  \label{tab:auc}
  \centering
  \begin{tabular}{l|r|r|r}
    \hline
    &           &     Quantum &    Quantum \\
    & Classical & (Simulator) & (Hardware) \\
    \hline
    \hline
    Sequence 1 & \textbf{0.994} & 0.988 & 0.981 \\ 
    Sequence 2 & 0.921 & 0.965 & \textbf{0.971} \\ 
    Sequence 3 & 0.973 & \textbf{0.997} & 0.996 \\ 
    Sequence 4 & 0.843 & \textbf{1.000} & \textbf{1.000} \\ 
    Sequence 5 & 0.957 & \textbf{0.993} & 0.991 \\ 
    Sequence 6 & 0.973 & \textbf{1.000} & \textbf{1.000} \\ 
    Sequence 7 & 0.940 & \textbf{1.000} & \textbf{1.000} \\ 
    \hline
  \end{tabular}
\end{table}

Next, we quantitatively evaluate the model performance.
The anomaly period can be viewed as labels for data points in a time series, hence we can define the ROC-AUC score for a labeled sequence.
This metric represents the performance of detection models taking the trade-off between the true positive and false positive into account without specifying the threshold.
Table~\ref{tab:auc} shows AUC scores on test sequences.
The quantum model outperforms the classical model on 6 out of 7 test instances.
The result indicates the potential of incorporating quantum feature transformation into traditional detection models.
Again, the hardware results are mostly similar to the simulator results, which suggests that application of near-term quantum computers to this task would be feasible enough.
Beyond evaluating the models using the AUC score, we further investigate the practical performance based on other metrics by setting the following threshold for the anomaly score.
We take the first seven anomaly scores after the normal period and set the threshold as a constant multiple of the mean of the scores.
This constant should be determined by trial operation on different data in practice, but here we choose a value that results in comparable false positive alert rates between our classical and quantum models, to enable explicit comparison between the models.
Specifically, the constant is set to 1.5 for the classical model and 3 for the quantum model.
Table~\ref{tab:false_alerts} shows the number of false alerts, i.e., the timestamps before the anomaly with the score exceeding the threshold.
The result of the time to detection for each model are shown in Table~\ref{tab:detection_time}.
We observe that the quantum model successfully detects the anomaly on all sequences, while keeping the number of the false alerts relatively small.
Overall, these results demonstrate the practical utility of the proposed quantum approach to failure detection.

\begin{table}[t]
  \caption{Number of False Alerts.}
  \label{tab:false_alerts}
  \centering
  \begin{tabular}{l|r|r|r}
    \hline
    &           &     Quantum &    Quantum \\
    & Classical & (Simulator) & (Hardware) \\
    \hline
    \hline
    Sequence 1 & 53 & 57 & 46 \\ 
    Sequence 2 & 14 & 12 & 12 \\ 
    Sequence 3 & 27 & 0 & 0 \\ 
    Sequence 4 & 80 & 29 & 68 \\ 
    Sequence 5 & 0 & 6 & 3 \\ 
    Sequence 6 & 36 & 20 & 11 \\ 
    Sequence 7 & 0 & 27 & 42 \\ 
    \hline
  \end{tabular}
\end{table}

\begin{table}[t]
  \caption{Detection Time for Machine Failure.}
  \label{tab:detection_time}
  \centering
  \begin{tabular}{r|r|r|r|r}
    \hline
     & Ground &           &     Quantum &    Quantum \\
     & Truth & Classical & (Simulator) & (Hardware) \\
    \hline
    \hline
    Seq. 1 & 2024-03-16 & 2024-03-16 & 2024-03-16 & 2024-03-17 \\ 
    Seq. 2 & 2024-01-13 & 2024-01-13 & 2024-01-13 & 2024-01-13 \\ 
    Seq. 3 & 2024-01-13 & 2024-01-13 & 2024-01-13 & 2024-01-13 \\ 
    Seq. 4 & 2024-03-10 & 2024-03-26 & 2024-03-26 & 2024-03-26 \\ 
    Seq. 5 & 2024-03-01 & Not Detected & 2024-03-01 & 2024-03-01 \\ 
    Seq. 6 & 2024-03-25 & 2024-03-27 & 2024-03-25 & 2024-03-25 \\ 
    Seq. 7 & 2024-03-25 & Not Detected & 2024-03-25 & 2024-03-25 \\ 
    \hline
  \end{tabular}
\end{table}

\section{Conclusion and Future Perspective}\label{sec:conclusion}

We introduced a machine failure detection framework based on a quantum machine learning approach.
We empirically validated the practical utility of projected quantum models for fault detection in a real-world scenario as well as on benchmark datasets.
This study’s contribution lies in the empirical exploration of quantum computing applications on industrial data using a real quantum device, paving the way for further investigations into harnessing the full potential of quantum computing in practice.
The exact reasons why projected quantum models yield precision improvements, particularly on noisy time series, remain unclear, posing intriguing challenges for future research.
In particular, behavior analysis of the quantum feature extraction from the perspective of Fourier analysis~\cite{schuld2021effect} for example would be an essential step toward understanding the impact on the performance of the detection model.


\appendix

\section{Additional Experiments}

\subsection{Experiments with Two-Local Circuit}\label{app:two-local}

\begin{figure}[t]
\centering
    \includegraphics[width=0.5\linewidth]{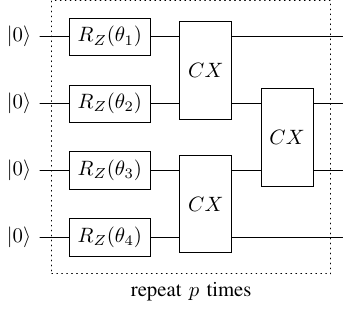}
    \caption{Schematic diagram of two-local quantum circuit for $n_q=4$ with parameters $\theta$.}
    \label{fig:two_local_circuit}
\end{figure}

\begin{figure}[t]
    \begin{subfigure}{.23\textwidth}
        \centering
        \includegraphics[width=0.95\linewidth]{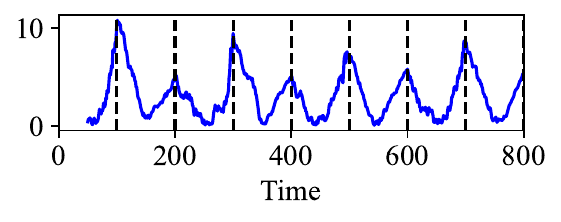}
        \subcaption{Synthetic data.}
    \end{subfigure}
    \begin{subfigure}{.23\textwidth}
        \centering
        \includegraphics[width=0.95\linewidth]{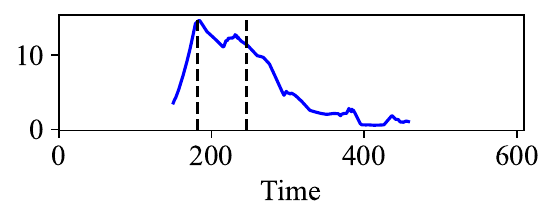}
        \subcaption{Bee waggle dance data.}
    \end{subfigure}
    \caption{Change-point scores with two-local ansatz.}
    \label{fig:two_local_score}
\end{figure}

In addition to the Heisenberg Hamiltonian circuit in the main text, we tested a circuit called the two-local circuit with alternating Z-rotation and CX gates as shown in Fig.~\ref{fig:two_local_circuit}.
We set the number of layer $p$ to 1.
We plot the change scores on change-detection experiments in Fig.~\ref{fig:two_local_score}.
We see similar results with those in the main text, while the second peak is slightly less emphasized on the bee waggle dance data.
Further exploration of quantum circuits suited for our application is left as future work.

\subsection{Experiments on Raw-resolution Dataset} \label{appendix:experiment_raw_dataset}

Table~\ref{tab:auc_raw} summarizes the AUC results on raw-resolution datasets without averaging over each day.
In this experiment, the number of the basis functions was set to $m=64$ and the regularization coefficient was $\lambda=0.1$ for uLSIF.
In comparison with the results on the daily data (Table~\ref{tab:auc}), the AUC scores slightly improves on most cases.
The reason is possibly that the statistical divergence estimation gets more accurate with the increased number of data points.
Nevertheless, the general trends are similar to the results on the daily data.
In particular, the quantum model (executed with a simulator) got better AUC scores than the classical model on 4 instances.

\begin{table}[t]
  \caption{AUC scores on raw resolution time series.}
  \label{tab:auc_raw}
  \centering
  \begin{tabular}{l|r|r}
    \hline
    &           &     Quantum  \\
    & Classical & (Simulator)  \\
    \hline
    \hline
    Sequence 1 & \textbf{0.998} & 0.988 \\ 
    Sequence 2 & 0.907 & \textbf{0.947} \\ 
    Sequence 3 & \textbf{1.000} & \textbf{1.000} \\ 
    Sequence 4 & \textbf{1.000} & \textbf{1.000} \\ 
    Sequence 5 & 0.970 & \textbf{1.000} \\ 
    Sequence 6 & 0.999 & \textbf{1.000} \\ 
    Sequence 7 & 0.878 & \textbf{1.000} \\ 
    \hline
  \end{tabular}
\end{table}


\bibliographystyle{IEEEtran}
\bibliography{ref}

\end{document}